# Growth, stabilization and conversion of semi-metallic and semiconducting phases of MoTe$_2$ monolayer by molecular-beam epitaxy

*Jinglei Chen, Guanyong Wang, Yanan Tang, Jinpeng Xu, Xianqi Dai, Jinfeng Jia, Wingkin Ho, and Maohai Xie[*]*

J. Chen, Dr. J. Xu, W. Ho, Prof. M. Xie

Physics Department

The University of Hong Kong

Pokfulam Road, Hong Kong

E-mail: mhxie@hku.hk

G. Wang, Prof. J. Jia

Key Laboratory of Artificial Structures and Quantum Control (Ministry of Education)

Department of Physics and Astronomy, Shanghai Jiaotong University

800 Dongchuan Road, Shanghai 200240, China;

Innovation Center of Advanced Microstructures, Nanjing 210093, China

E-mail: jfjia@sjtu.edu.cn

Y. Tang, Prof. X. Dai

College of Physics and Electronic Engineering

Henan Normal University

Xinxiang, Henan 453007, China;

School of Physics and Electronic Engineering

Zhengzhou Normal University

Henan 450044, China

E-mail: xqdaizs@163.com

Transition-metal dichalcogenides (TMDs) with the common formula $MX_2$ (M = Mo, W; X=S, Se, Te) exist in different phases such as the hexagonal (2H), octahedral (1T), monoclinic (1T') and orthorhombic (Td) structures.[1, 2, 3-7] The 2H phase is most common including, for example, metal disulfides ($MS_2$) and diselenides ($MSe_2$), which are direct gap semiconductors for monolayer (ML) thin films.[1, 8] They have attracted extensive research attention in recent years due to their appeals in microelectronic, optoelectronic, spin and valley electronic applications.[1, 9, 10, 11, 12] The 1T' or Td phase $MX_2$ are of the distorted 1T structure, which usually show semi-metallic behavior.[3, 13-15] Examples of the latter include $WTe_2$ and $β$-$MoTe_2$, which have drawn special interests lately following some recent revelations of, e.g., the large and unsaturated magnetoresistance,[4, 15, 16, 17, 18] pressure-driven superconductivity,[7, 19] novel optical properties and characteristics,[11, 12, 17, 18, 20] and the topological insulator[14] and Weyl semimetal states.[6, 7, 21] Metallic TMDs are also good catalysts for hydrodesulfurization and hydrogen evolution reactions.[22] The large difference in electrical properties between 2H and 1T' (Td) phases of $MX_2$ further makes them promising for phase-change electronics.[23, 24] Therefore, tuning and stabilizing the different phases of $MX_2$ can be of great scientific and application relevance.

Among the various TMDs, $MoTe_2$ takes a special place as there is a small energy difference between its 2H and 1T' phases (~ 43meV per formula unit[5]). The hexagonal phase of $MoTe_2$ is slightly more stable than 1T' $MoTe_2$ under ambient conditions, while the latter becomes more favorable at high temperature and/or under tensile strain.[5, 24] In any case, due to the small energy difference between the two structures, there is a high chance for one to obtain samples containing coexisting phases or to purposely tune the structure of MoTe$_2$ crystal by applying external constraints. This would lead to many new applications of the TMD thin films.[18]

In this work, we report growths of both 2H and 1T' MoTe$_2$ ML by molecular beam epitaxy (MBE). We reveal a dramatic effect of Te adsorption on 1T' phase MoTe$_2$ formation and growth. By changing the conditions of MBE and by annealing, we can achieve effective tuning of the structural phase of MoTe$_2$. Employing scanning tunneling microscopy and spectroscopy (STM/S), we establish unambiguously the structures and electronic characteristics of 2H and 1T' MoTe$_2$ ML such as the energy bandgap and the density of states (DOS). By consulting with the first principles calculations, we provide an explanation for the stabilization of the otherwise metastable 1T' phase MoTe$_2$ at the temperature and pressure condition, which is associated with Te adsorption on surface.

**Figure 1**a shows a STM topographic image of an as-grown MoTe$_2$ epilayer deposited at 250 ℃ on highly ordered pyrolytic graphite (HOPG) for a nominal coverage of 0.5 MLs. From it, one discerns clearly monolayer high islands as confirmed by line profile measurement (Figure 1b). Reflection high-energy electron diffraction (RHEED) pattern of the surface is streaky (Figure 1c), implying a flat morphology of the sample. A close look of the RHEED pattern reveals, however, that there are two sets of diffraction streaks as marked by the thin arrows in Figure 1c. By measuring the inter-streak spacing, which represent the reciprocals of surface lattice

constants, the two sets of streaks can be consistently assigned to the 2H and 1T' phase MoTe$_2$, i.e., one corresponds to $b^{1T'} \sim 6.25$ Å of the 1T' phase MoTe$_2$ while the other to $\sqrt{3}a^{2H} \sim 6.08$ Å of the 2H MoTe$_2$. Such epifilms containing coexisting phases of 2H and 1T' MoTe$_2$ are found quite common for growth at low temperatures and using high Te fluxes. As will be discussed later, we may associate the 1T' phase MoTe$_2$ formation and growth with Te adsorption, so changing the growth conditions leads to variations of surface Te coverage, which in turn changes the relative domain size of 1T' versus 2H MoTe$_2$ in the epifilm.

The coexistence of 2H and 1T' phase MoTe$_2$ in the epifilm is confirmed by close-up STM measurements as presented in Figs. 2 and 3 below. The 2H phase has the hexagonal symmetry of the surface lattice (**Figure 2**a). The close-up STM image of such a domain given in Figure 2b reveals a dense triangular network of the bright lines, which resembles that in epitaxial MoSe$_2$ grown by MBE.[25] These bright lines in MoSe$_2$ epifilms were identified to be inversion domain boundary defects, and the triangular network of the lines reflects the very symmetry of 2H MX$_2$ lattice. STS measurements at the center of the triangle (i.e., away from the lines) reveal a DOS gap, as seen in Figure 2c, confirming that it is a semiconductor. This experimental spectrum agrees with the density functional theory (DFT) calculated DOS of 2H MoTe$_2$ ML shown in Figure 2d. From the experimental STS data, we further derive an electronic energy gap of ~ 1.4 eV (refer to Supplementary), a value that is in qualitative agreement with that reported in literature by optical studies. The latter can, however, be offset by an amount equal to the binding energy of excitons in the

material.[10, 12, 26]

The 1T' phase MoTe$_2$ has the monoclinic lattice structure as shown in **Figure 3**a. Figure 3b is an atomic-resolution STM image of such a domain in sample, from which one notes clearly the rectangular structure of the surface atoms being consistent with 1T' phase MoTe$_2$ ML. Lattice constants of $a \sim 3.52$ Å and $b \sim 6.25$ Å can be measured directly from the image, which differ slightly from that of a free-standing MoTe$_2$ ML (i.e., $a \sim 3.5$ Å, $b \sim 6.3$ Å). The STM measured lattice constant agrees with the RHEED measurement cited above and the slight distortion with respect to the free-standing ML is likely caused by HOPG substrate. STS measurements over such domains reveal no DOS gap, which is consistent with the semi-metallic MoTe$_2$ and agrees well with the theoretically calculated DOS as presented in Figure 3d.

With the finding that both 2H and 1T' domains exist in one and the same MoTe$_2$ ML sample, an important issue is what derives the metastable phase to grow and stabilize and whether the phases can be tuned by changing growth conditions. Obviously, the coexisting phases of MoTe$_2$ is connected to the small energy difference between 2H and 1T' phases of MoTe$_2$. For MBE growth, however, kinetics may also play a role in stabilizing the otherwise metastable phase at the experimental conditions. Varying the temperature and source flux has indeed been found to change the relative size of the domains, where the lower the temperature and the higher the Te flux, the larger the 1T' domain in the epifilm. At high temperatures ($> 400$ °C) and low Te flux ($< 0.5 \times 10^{13}$ molecules/cm$^2$·s), 2H phase would become predominant with little trace of 1T'-MoTe$_2$ discerned by the RHEED and STM measurements.

To quantify the relative domain size of the 1T' versus 2H MoTe$_2$ in the epifilm, we extract the intensities ($I$) of the two sets of the RHEED streaks by least-square multi-peak fittings (refer to Supplementary). The intensity ratio $R = I_{1T'}/(I_{1T'} + I_{2H})$ is calculated and plotted in Figure 4a against the (inverse) growth temperature $T$. To the first-order approximation, we may correlate the intensity ratio $R$ with 1T' domain area, which, according to **Fig. 4**a, follows an Arrhenius relation: $R \propto e^{E/k_B T}$, where $k_B$ is the Boltzmann constant and $E$ is the energy of activation. Lowering the temperature leads to an increasing $R$. The same is found by increasing Te flux (*c.f.*, the open circle in Figure 4a. See also Figure S3 in the Supplementary). Based on these results, we suggest that the nucleation and growth of the 1T'-phase MoTe$_2$ is likely promoted by Te adsorbate on surface. Lowering the temperature and increasing the source flux can both lead to increasing Te coverage, which have the same effect of enlarging 1T' MoTe$_2$ domain area in the sample.

According to first-order adsorption and desorption kinetics, the surface coverage $\theta$ of the adsorbate increases, in the low coverage limit, at the rate of

$$\frac{d\theta}{dt} = k_a F - k_d \theta \qquad (1)$$

where $F$ stands for the deposition flux, $k_a$ and $k_d$ are the rate constants of adsorption and desorption processes, respectively. In the above, we have neglected the effect of atom incorporation in film due to the low growth rate of MoTe$_2$. At steady state, $\frac{d\theta}{dt} = 0$ and so $k_a F = k_d \theta$, from which one gets

$$\theta \sim \frac{F}{K} \propto e^{E/k_B T} \text{ with } K = \frac{k_d}{k_a} \propto e^{-E/k_B T} \qquad (2)$$

As we suggest above, if $R \propto \theta$, the data in Figure 4a allow us to derive an energy according to eqn. (2), which is approximately 0.17 eV. This would reflect the difference between energies of desorption and adsorption processes.

The suggestion that the 1T'-phase MoTe$_2$ is favored at high Te coverage may be further tested by annealing experiment. The results are given in Supplementary S4. As one may note, increasing the temperature of annealing while keeping the annealing time constant, the 1T' domain size ($R$) decreases monotonically. On the other hand, if we perform the same annealing experiments but under a Te flux, the rate of decrease of $R$ becomes greatly suppressed, in strong support of above suggestion of Te adsorbate effect in promoting 1T' MoTe$_2$ growth.

Finally, to further elucidate the role of Te adsorbate on 1T'-MoTe$_2$ formation and stability, we have performed first-principles total energy calculations of systems consisted of MoTe$_2$ MLs laid on top of graphene with varying numbers of Te atom(s) adsorbed on MoTe$_2$ surface. The graphene layer models the HOPG substrate used in experiment. We firstly compare free-standing 2H and 1T' MoTe$_2$ ML and find indeed that the 2H phase had a lower energy than 1T', in agreement with the literature. On graphene, however, our calculation shows a reduction in the energy difference between the two phases, implying that the substrate helps to stabilize the 1T' MoTe$_2$. Even more dramatic effect is found by Te adsorption. Specifically, we compare the total energies of the 1T' versus 2H phases MoTe$_2$ on graphene with increasing numbers of Te adatoms on MoTe$_2$ surface. By comparing different adsorption configurations, we identify that the most favorable adsorption site is the hollow site but above the Mo atom (refer to Supplementary S5). Figure 4b summarizes the total energies of systems of (4 × 4) MoTe$_2$ on (6 × 6) graphene 'supercell' with 0 – 8 Te adatoms. The red solid and black open symbols represent the 2H and 1T' MoTe$_2$, respectively. All energies are measured relative to the clean 2H MoTe$_2$ without adsorbate. As can be seen, when the number of Te adatoms (equivalently Te coverage) increases, the 2H phase shows an increasing trend in energy, implying that 2H MoTe$_2$ is increasingly unstable. In contrast, the 1T' phase shows a decreasing energy with increasing Te coverage, suggesting that 1T' MoTe$_2$ becomes increasingly favored by Te adsorption. This is in accord with the above experimental finding.

To conclude, both 2H and 1T' phase MoTe$_2$ have been grown by MBE on HOPG substrate. STM measurements reveal the 2H domains to contain high density of line defects arranged in triangular networks, which resemble that of epitaxial MoSe$_2$. STS measurements over such 2H MoTe$_2$ domains reveal an electronic gap of ~1.4 eV, which is in agreement with theory and optical measurements. Atomic-resolution STM images of the 1T' phase MoTe$_2$ clearly resolve the monoclinic crystal structure but slightly strained lattice, presumably by the effect of the substrate. STS measurements confirm the semi-metallic behavior of the 1T' MoTe$_2$. Interestingly, we find the domain size of the 1T' MoTe$_2$ can be tuned effectively by changing the deposition conditions of MBE, which reflects an important effect of Te adsorption on structural phase stabilization of epitaxial MoTe$_2$. First principles calculations support the experimental finding. The result suggests an effective and convenient method to tune the structural phases of MoTe$_2$ during MBE, which provides a true opportunity to exploring phase-change electronics based on TMDs.

**Experimental Section**

*MBE Growth*: deposition of MoTe$_2$ on HOPG using elemental Mo and Te sources was carried out in a customized Omicron MBE reactor having the based pressure of 5 × 10$^{-10}$ mbar. The fluxes of Mo and Te were generated from an e-beam cell and a conventional Knudsen cell, respectively, and the latter (Te) was set a few tens to hundred times that of Mo to ensure the stoichiometry and crystallinity of epitaxial MoTe$_2$. Film growth rate was about 0.3 MLs h$^{-1}$, which was limited by Mo flux and determined by post-growth coverage measurements of the deposit by STM. The growth temperature ranged between 250 ℃ and 400 ℃. Prior to MoTe$_2$ deposition, the HOPG substrate was thoroughly degassed in the UHV chamber and flashed up to 600 ℃. During film deposition, the surface was constantly monitored by the RHEED operated at 15 keV. Annealing of samples at varying temperatures were done in the MBE chamber with or without the supply of Te flux.

*STM/S Measurements*: morphological and spectroscopic measurements of the grown samples were carried out using either a room-temperature Omicron STM system connected to the MBE chamber or a low-temperature Unisoku STM facility ex situ. For the latter, the sample was protected by depositing an amorphous Te layer at RT before being taken out of the vacuum. After loaded into the Unisoku system, the sample underwent a gentle annealing process (~250 ℃ for 1hr) to desorb the Te capping layer, which could be confirmed both by the recovery of the streaky RHEED pattern and by STM examination of the surface showing the same terrace-and-step morphology. Topographic imaging by STM and the STS measurements were carried out at 77 K. For the latter, lock-in technique was used with the modulation frequency of 991 Hz and voltage of 5 mV. Each spectrum presented in the paper represents an average of 30 measurements at the same position.

*First principles calculations*: all calculations, including spin-polarized DFT calculations, were performed using the Vienna ab initio simulation package (VASP).[27] To improve the calculation efficiency, core electrons were replaced by the projector augmented wave (PAW) pseudo-potentials.[28] Both the van der Waals density functional (vdW-DF)[29] and the Perdew, Burke, and Ernzernhof (PBE)[30] were adopted to describe the effects of dispersive interactions between $MoTe_2$ and graphene substrate. The Kohn-Sham orbitals were expanded using plane waves expansion with an energy cutoff of 450 eV. The convergence criterion for electronic self-consistent iteration was set to $10^{-5}$ eV. The C 2s2p, Mo 4d5s and Te 4d5s5p states were treated as valence electrons. In the calculation, a 15×15×1 k-points grid was used. The energy of an isolated atom was simulated using a cubic cell of 15×15×15 $Å^3$ with one atom putting in the middle. To simulate ML $MoTe_2$ on graphene substrate, we chose optimized (4×4) unit cells of $MoTe_2$ laid on top of (6×6)-graphene and the vacuum size was larger than 15 Å.

**Acknowledgements**

The work described in this paper was supported by a Collaborative Research Fund (HKU9/CRF/13G) sponsored by the Research Grant Council (RGC), Hong Kong Special Administrative Region. J.J. wishes to thank the Ministry of Science and Technology of China (2013CB921902), NSFC (11521404, 11227404) for partial support. X.D. acknowledges NSFC (11504334 and U1404109) for support.

Received:
Revised:
Published online:

# References

[1] G. B. Liu, D. Xiao, Y. Yao, X. Xu, W. Yao, Chemical Society reviews 2014.

[2] B. Brown, Acta Crystallographica 1966, 20, 268; S. L. Tang, R. V. Kasowski, B. A. Parkinson, Physical Review B 1989, 39, 9987.

[3] W. G. Dawson, D. W. Bullett, Journal of Physics C: Solid State Physics 1987, 20, 6159.

[4] M. N. Ali, J. Xiong, S. Flynn, J. Tao, Q. D. Gibson, L. M. Schoop, T. Liang, N. Haldolaarachchige, M. Hirschberger, N. P. Ong, R. J. Cava, Nature 2014, 514, 205.

[5] K. A. Duerloo, Y. Li, E. J. Reed, Nature communications 2014, 5, 4214.

[6] A. A. Soluyanov, D. Gresch, Z. Wang, Q. Wu, M. Troyer, X. Dai, B. A. Bernevig, Nature 2015, 527, 495.

[7] Y. Qi, P. G. Naumov, M. N. Ali, C. R. Rajamathi, W. Schnelle, O. Barkalov, M. Hanfland, S. C. Wu, C. Shekhar, Y. Sun, V. Suss, M. Schmidt, U. Schwarz, E. Pippel, P. Werner, R. Hillebrand, T. Forster, E. Kampert, S. Parkin, R. J. Cava, C. Felser, B. Yan, S. A. Medvedev, Nature communications 2016, 7, 11038.

[8] Q. H. Wang, K. Kalantar-Zadeh, A. Kis, J. N. Coleman, M. S. Strano, Nature nanotechnology 2012, 7, 699; Y. Zhang, T. R. Chang, B. Zhou, Y. T. Cui, H. Yan, Z. Liu, F. Schmitt, J. Lee, R. Moore, Y. Chen, H. Lin, H. T. Jeng, S. K. Mo, Z. Hussain, A. Bansil, Z. X. Shen, Nature nanotechnology 2014, 9, 111; K. F. Mak, C. Lee, J. Hone, J. Shan, T. F. Heinz, Physical Review Letters 2010, 105.

[9] D. Xiao, G.-B. Liu, W. Feng, X. Xu, W. Yao, Physical Review Letters 2012, 108; T. Cao, G. Wang, W. Han, H. Ye, C. Zhu, J. Shi, Q. Niu, P. Tan, E. Wang, B. Liu, J. Feng, Nature communications 2012, 3, 887; K. F. Mak, K. He, J. Shan, T. F. Heinz, Nature nanotechnology 2012, 7, 494; H. Zeng, J. Dai, W. Yao, D. Xiao, X. Cui, Nature nanotechnology 2012, 7, 490; B. Radisavljevic, A. Radenovic, J. Brivio, V. Giacometti, A. Kis, Nature nanotechnology 2011, 6, 147; K. Roy, M. Padmanabhan, S. Goswami, T. P. Sai, G. Ramalingam, S. Raghavan, A. Ghosh, Nature nanotechnology 2013, 8, 826; C. Chakraborty, L. Kinnischtzke, K. M. Goodfellow, R. Beams, A. N. Vamivakas, Nature nanotechnology 2015; T. Böker, R. Severin, A. Müller, C. Janowitz, R. Manzke, D. Voß, P. Krüger, A. Mazur, J. Pollmann, Physical Review B 2001, 64; Y. Ma, Y. Dai, M. Guo, C. Niu, J. Lu, B. Huang, Physical chemistry chemical physics : PCCP 2011, 13, 15546; H. Guo, T. Yang, M. Yamamoto, L. Zhou, R. Ishikawa, K. Ueno, K. Tsukagoshi, Z. Zhang, M. S. Dresselhaus, R. Saito, Physical Review B 2015, 91; J. Park, Y. Kim, Y. I. Jhon, Y. M. Jhon, Applied Physics Letters 2015, 107, 153106; M. Kuiri, B. Chakraborty, A. Paul, S. Das, A. K. Sood, A. Das, Applied Physics Letters 2016, 108, 063506; L. Yin, X. Zhan, K. Xu, F. Wang, Z. Wang, Y. Huang, Q. Wang, C. Jiang, J. He, Applied Physics Letters 2016, 108, 043503; H. Zhang, W. Zhou, X. Li, J. Xu, Y. Shi, B. Wang, F. Miao, Applied Physics Letters 2016, 108, 091902.

[10] I. G. Lezama, A. Ubaldini, M. Longobardi, E. Giannini, C. Renner, A. B. Kuzmenko, A. F. Morpurgo, 2D Materials 2014, 1, 021002; C. Ruppert, O. B. Aslan, T. F. Heinz, Nano letters 2014, 14, 6231.

[11] G. Froehlicher, E. Lorchat, F. Fernique, C. Joshi, A. Molina-Sanchez, L. Wirtz, S. Berciaud, Nano letters 2015, 15, 6481.

[12] S. Koirala, S. Mouri, Y. Miyauchi, K. Matsuda, Physical Review B 2016, 93.

[13] V. E. J. Augustin, Physical Review B 2000, 62.

[14] J. L. Xiaofeng Qian, Liang Fu, Ju Li, Science 2014, 346.

[15] I. Pletikosić, M. N. Ali, A. V. Fedorov, R. J. Cava, T. Valla, Physical Review Letters 2014, 113.

[16] J. Jiang, F. Tang, X. C. Pan, H. M. Liu, X. H. Niu, Y. X. Wang, D. F. Xu, H. F. Yang, B. P. Xie, F. Q. Song, P. Dudin, T. K. Kim, M. Hoesch, P. K. Das, I. Vobornik, X. G. Wan, D. L. Feng, Physical Review Letters

2015, 115; H. Y. Lv, W. J. Lu, D. F. Shao, Y. Liu, S. G. Tan, Y. P. Sun, EPL (Europhysics Letters) 2015, 110, 37004; D. Rhodes, S. Das, Q. R. Zhang, B. Zeng, N. R. Pradhan, N. Kikugawa, E. Manousakis, L. Balicas, Physical Review B 2015, 92; Y. L. Wang, L. R. Thoutam, Z. L. Xiao, J. Hu, S. Das, Z. Q. Mao, J. Wei, R. Divan, A. Luican-Mayer, G. W. Crabtree, W. K. Kwok, Physical Review B 2015, 92; P. K. Das, D. Di Sante, I. Vobornik, J. Fujii, T. Okuda, E. Bruyer, A. Gyenis, B. E. Feldman, J. Tao, R. Ciancio, G. Rossi, M. N. Ali, S. Picozzi, A. Yadzani, G. Panaccione, R. J. Cava, Nature communications 2016, 7, 10847.

[17] Y. Luo, H. Li, Y. M. Dai, H. Miao, Y. G. Shi, H. Ding, A. J. Taylor, D. A. Yarotski, R. P. Prasankumar, J. D. Thompson, Applied Physics Letters 2015, 107, 182411.

[18] D. H. Keum, S. Cho, J. H. Kim, D.-H. Choe, H.-J. Sung, M. Kan, H. Kang, J.-Y. Hwang, S. W. Kim, H. Yang, K. J. Chang, Y. H. Lee, Nature Physics 2015, 11, 482.

[19] X. C. Pan, X. Chen, H. Liu, Y. Feng, Z. Wei, Y. Zhou, Z. Chi, L. Pi, F. Yen, F. Song, X. Wan, Z. Yang, B. Wang, G. Wang, Y. Zhang, Nature communications 2015, 6, 7805; D. Kang, Y. Zhou, W. Yi, C. Yang, J. Guo, Y. Shi, S. Zhang, Z. Wang, C. Zhang, S. Jiang, A. Li, K. Yang, Q. Wu, G. Zhang, L. Sun, Z. Zhao, Nature communications 2015, 6, 7804.

[20] C. C. Homes, M. N. Ali, R. J. Cava, Physical Review B 2015, 92; W. D. Kong, S. F. Wu, P. Richard, C. S. Lian, J. T. Wang, C. L. Yang, Y. G. Shi, H. Ding, Applied Physics Letters 2015, 106, 081906.

[21] T.-R. Chang, S.-Y. Xu, G. Chang, C.-C. Lee, S.-M. Huang, B. Wang, G. Bian, H. Zheng, D. S. Sanchez, I. Belopolski, N. Alidoust, M. Neupane, A. Bansil, H.-T. Jeng, H. Lin, M. Zahid Hasan, Nature communications 2016, 7, 10639.

[22] D. Voiry, H. Yamaguchi, J. Li, R. Silva, D. C. Alves, T. Fujita, M. Chen, T. Asefa, V. B. Shenoy, G. Eda, M. Chhowalla, Nature materials 2013, 12, 850; D. Voiry, M. Salehi, R. Silva, T. Fujita, M. Chen, T. Asefa, V. B. Shenoy, G. Eda, M. Chhowalla, Nano letters 2013, 13, 6222; M. A. Lukowski, A. S. Daniel, F. Meng, A. Forticaux, L. Li, S. Jin, Journal of the American Chemical Society 2013, 135, 10274.

[23] T. F. Goki Eda, Hisato Yamaguchi, Damien Voiry, Mingwei Chen, Manish Chhowalla, ACS nano 2012, 6, 7311; S. Cho, S. Kim, J. H. Kim, J. Zhao, J. Seok, D. H. Keum, J. Baik, D. H. Choe, K. J. Chang, K. Suenaga, S. W. Kim, Y. H. Lee, H. Yang, Science 2015, 349, 625; Y. Li, K. A. Duerloo, K. Wauson, E. J. Reed, Nature communications 2016, 7, 10671.

[24] S. Song, D. H. Keum, S. Cho, D. Perello, Y. Kim, Y. H. Lee, Nano letters 2016, 16, 188.

[25] H. Liu, L. Jiao, F. Yang, Y. Cai, X. Wu, W. Ho, C. Gao, J. Jia, N. Wang, H. Fan, W. Yao, M. Xie, Phys Rev Lett 2014, 113, 066105.

[26] B. Chen, H. Sahin, A. Suslu, L. Ding, M. I. Bertoni, F. M. Peeters, S. Tongay, ACS nano 2015, 9, 5326.

[27] G. Kresse, J. Furthmuller, Physical Review B 1996, 54, 11169.

[28] G. Kresse, D. Joubert, Physical Review B 1999, 59, 1758.

[29] M. Dion, H. Rydberg, E. Schroder, D. C. Langreth, B. I. Lundqvist, Phys Rev Lett 2004, 92, 246401; T. Thonhauser, V. R. Cooper, S. Li, A. Puzder, P. Hyldgaard, D. C. Langreth, Physical Review B 2007, 76.

[30] K. B. John P. Perdew, Matthias Ernzerhof, Phys Rev Lett 1996, 77, 3865.

**Figures and Captions**

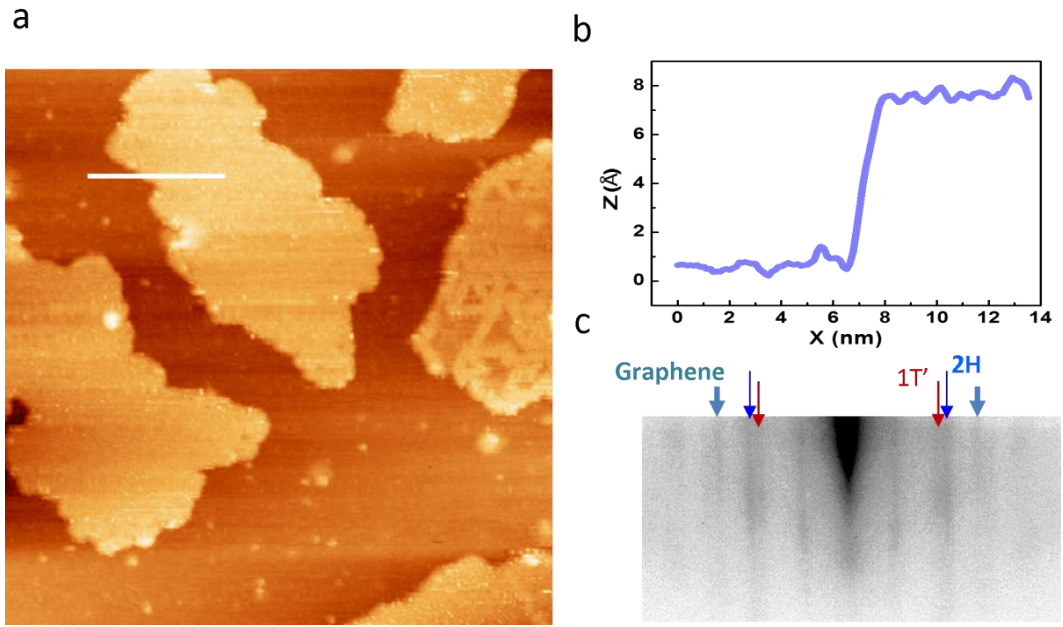

**Fig. 1 STM and RHEED:** **(a)** STM topographic image (size: $50 \times 50$ nm$^2$, sample bias: 0.5 V) of an as-grown MoTe$_2$ sample showing ML MoTe$_2$ islands/domains on the HOPG substrate. **(b)** Line profile of the surface along the white line in (a). **(c)** RHEED pattern of the sample showing two sets of diffraction streaks corresponding to the 2H and 1T' MoTe$_2$, respectively, as marked by the thin arrows. The pattern from the substrate surface is also indicated by short thick arrows.

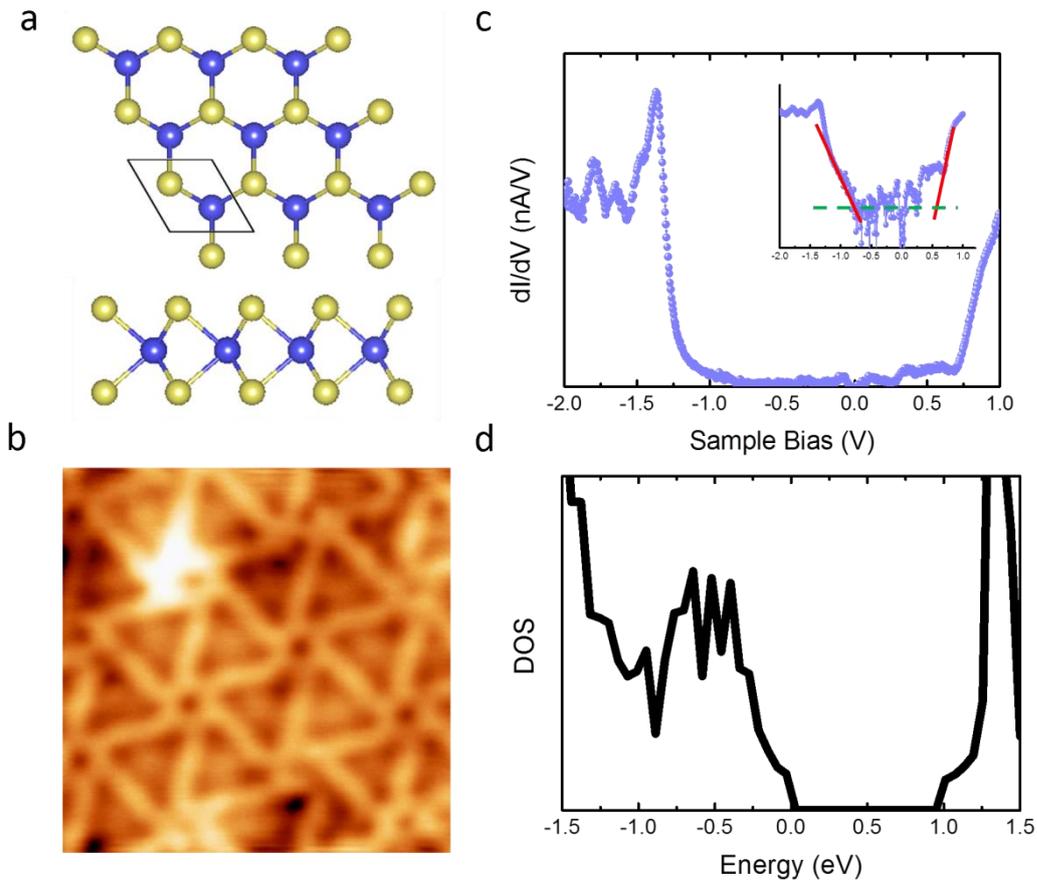

**Fig. 2 2H-MoTe$_2$:** **(a)** Stick-and-ball model of 2H MoTe2 ML viewed from top (upper) and side (lower). The blue and yellow balls represent Mo and Te atoms, respectively. **(b)** Close-up STM image (Size: 8 × 8 nm$^2$, sample bias: 2 V) of the 2H MoTe$_2$ domain. The triangular features reflect domain boundary defects, which resemble that in epitaxial MoSe$_2$.[25] **(c)** STS of 2H-MoTe$_2$ ML taken at the center of the triangle in (b). The inset shows the same spectrum but plotted in semi-logarithm scale, revealing the DOS gap of ~1.4 eV. **(d)** DFT calculated DOS of 2H-MoTe$_2$ ML.

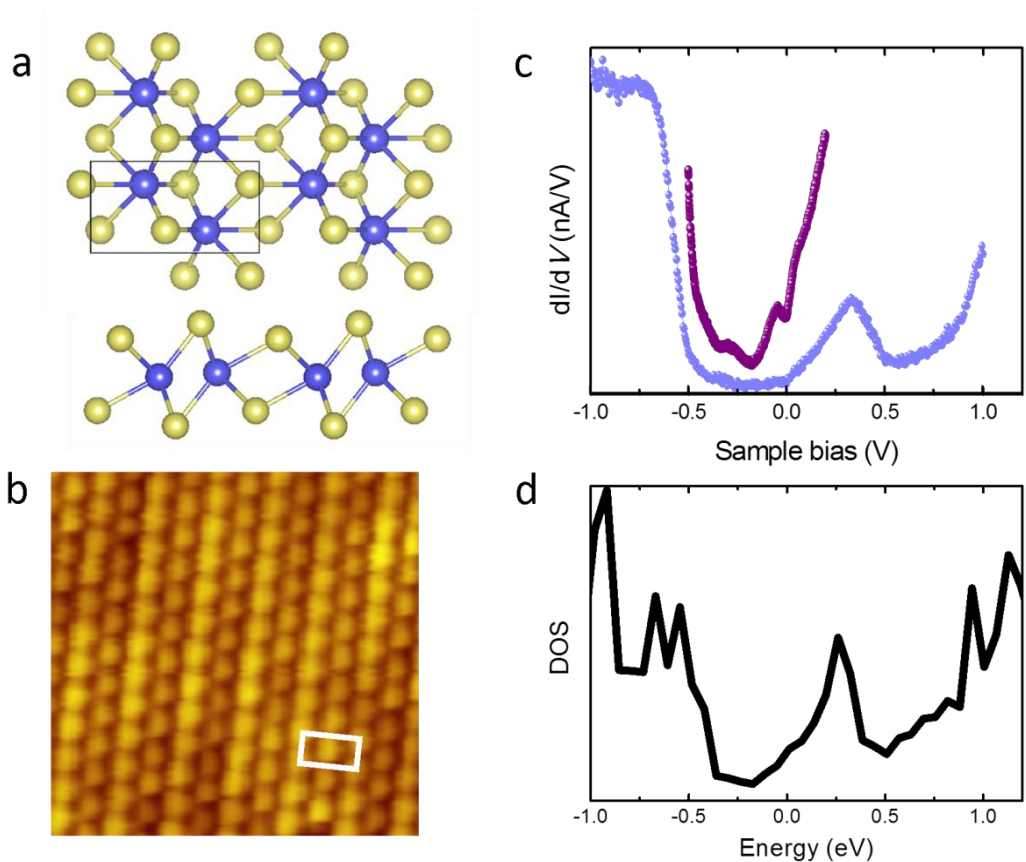

**Fig. 3 1T'-MoTe$_2$:** **(a)** Stick-and-ball model of 1T' MoTe2 ML viewed from top (upper) and side (lower). The blue and yellow balls represent Mo and Te atoms, respectively. **(b)** Atomic resolution STM image (Size: $4 \times 4$ nm$^2$, sample bias: +4.88 mV) of 1T' MoTe$_2$ domain in sample. **(c)** STS of 1T'-MoTe$_2$ ML, where the lower (blue) and upper (red) curves are measured over different energy ranges but otherwise at the same position. **(d)** DFT calculated DOS of 1T'-MoTe$_2$ ML.

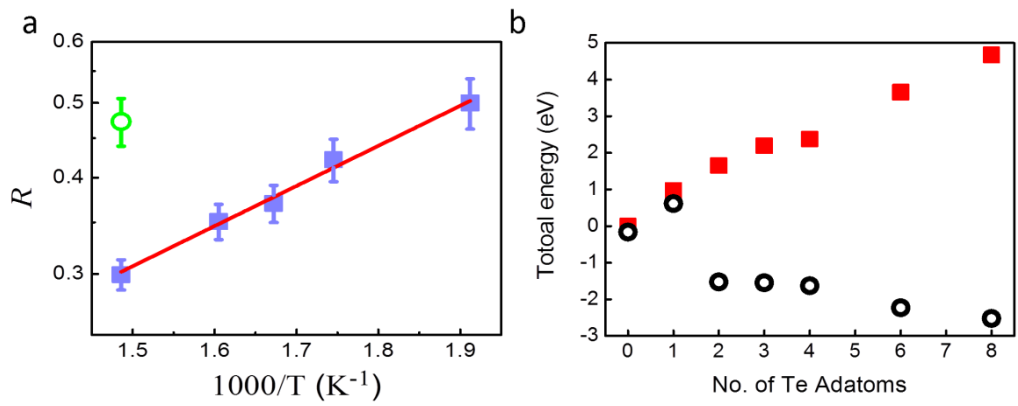

**Fig. 4 Phase domain tuning: (a)** Intensity ratio ($R$) of the RHEED streaks plotted as a function of (inverse) deposition temperature ($T$). The solid squares represent data obtained at the Te flux of $3.4 \times 10^{13}$ molecules/cm$^2$·s while the open circle is for growth at a higher Te flux of $1.0 \times 10^{14}$ molecules/cm$^2$·s. **(b)** DFT calculated total energies of 1T' (black circles) and 2H (red squares) MoTe$_2$-on-graphene with varying numbers of Te adatoms per (4 × 4) MoTe$_2$ supercell.

# Supplementary information

**Growth, stabilization and conversion of semi-metallic and semiconducting phases of MoTe$_2$ monolayer by molecular-beam epitaxy**

*Jinglei Chen, Guanyong Wang, Yanan Tang, Jinpeng Xu, Xianqi Dai, Jinfeng Jia, Wingkin Ho, and Maohai Xie[*]*

J. Chen, Dr. J. Xu, W. Ho, Prof. M. Xie

Physics Department

The University of Hong Kong

Pokfulam Road, Hong Kong

E-mail: mhxie@hku.hk

G. Wang, Prof. J. Jia

Key Laboratory of Artificial Structures and Quantum Control (Ministry of Education)

Department of Physics and Astronomy, Shanghai Jiaotong University

800 Dongchuan Road, Shanghai 200240, China;

Innovation Center of Advanced Microstructures, Nanjing 210093, China

E-mail: jfjia@sjtu.edu.cn

Y. Tang, Prof. X. Dai

College of Physics and Electronic Engineering

Henan Normal University

Xinxiang, Henan 453007, China;

School of Physics and Electronic Engineering

Zhengzhou Normal University

Henan 450044, China

E-mail: xqdaizs@163.com

**S1  Bandgap determination**

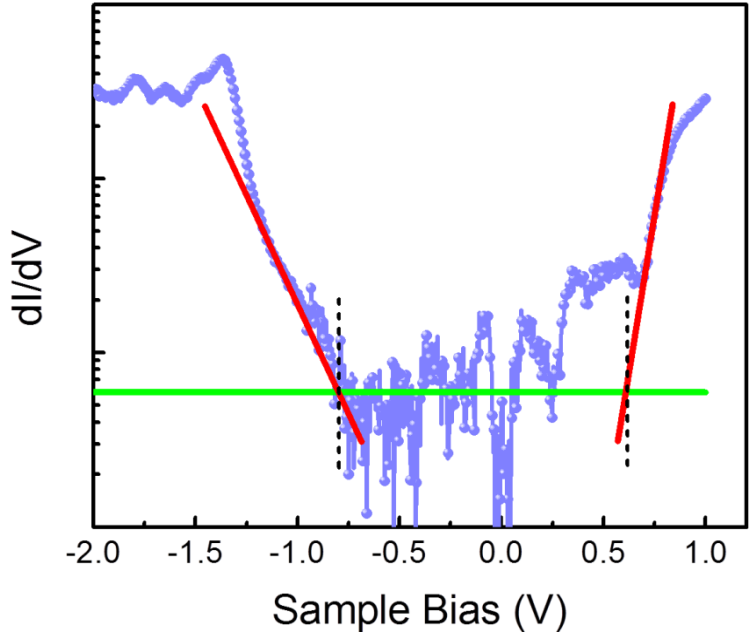

**Figure S1: Bandgap determination from STS curve**

We follow the method of M. F. Crommie[S1] to estimate the energy bandgap of the 2H-phase MoTe$_2$ ML based on the experimental STS spectrum as shown in **Figure S1**. The spectrum obtained at 77 K is plotted in logarithm scale. The electronic gap is determined by finding the width of the zero-conductance floor where the band edges are assigned by the intersections of the zero-conductance floor with the linear fits of the conductance data in regions of $E_{VB,2\sigma} - \Delta E < E < E_{VB,2\sigma}$ for the valence band edge and $E_{CB,2\sigma} < E < E_{CB,2\sigma} + \Delta E$ for the conduction band edge as shown by the red lines in Figure S1. We have chosen ΔE = 50 mV throughout. The complication and uncertainty here arise from the effect of the triangular network of line defects in such films. Because the triangles are relatively small, defects states as well as quantum confinement may affect the measurement, giving rise to non-zero DOS even in the gap region. So there is an uncertainty in determining the zero-conductance floor. Here we have arbitrarily taken it to be the average between -0.7 eV to -0.2 eV, and the estimated bandgap is 1.4 eV, which likely represents an upper bound.

## S2  Extracting the RHEED streak intensities

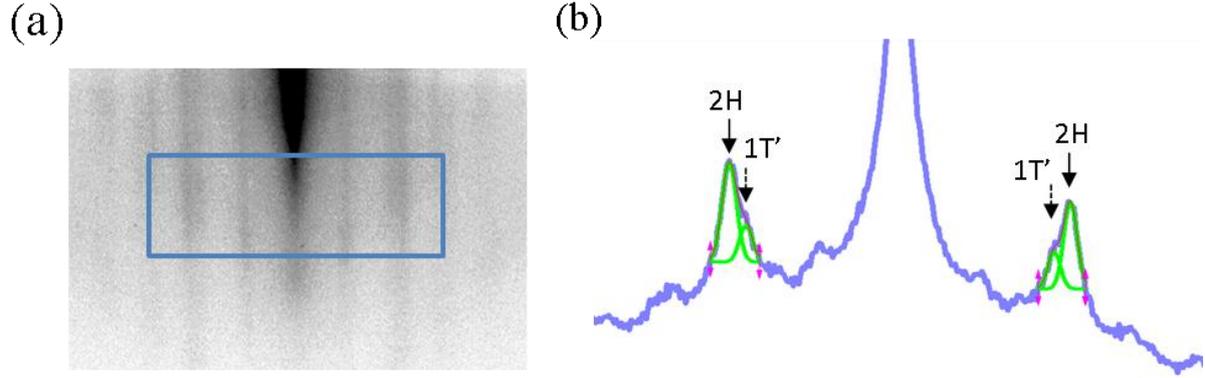

**Figure S2: (a) RHEED pattern of the sample surface. (b) Averaged intensity profile over the region of the rectangle box in (a)**

To extract the intensities of the RHEED streaks, we firstly take average along the vertical over the selected rectangular box as shown in **Figure S2**a and plotted as a function of the horizontal position in Fig. S2b. Then the two closely spaced satellite peaks corresponding to diffraction from 2H and 1T' MoTe$_2$, respectively, are resolved by multi-peak fittings (green lines in Fig. S2b) and the fitted intensities $I_{1T'}$ and $I_{2H}$ are readout to find the ratio $R = I_{1T'}/(I_{1T'} + I_{2H})$. The data presented in the Fig. 4a of the main text represent averages of three independent measurements and fitting procedures.

## S3 Experiments under different Te fluxes

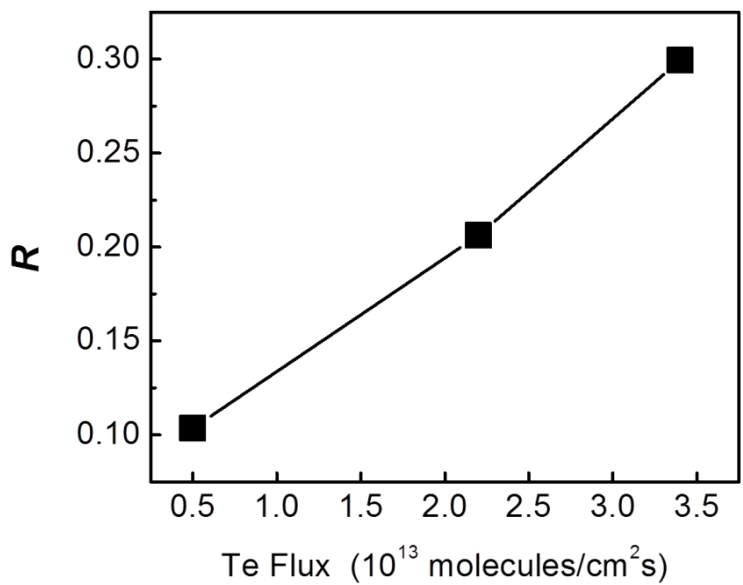

**Figure S3. The RHEED intensity ratio *R* vs. Te flux (see main text).**

Growth using different Te fluxes but at the same temperature is shown in **Figure S3**, where the growth temperature was constant at 400 ℃. It can be seen that the RHEED intensity ratio *R*, which is related to the domain size of the 1T' $MoTe_2$ increases with Te flux.

## S4  Annealing with and without Te flux

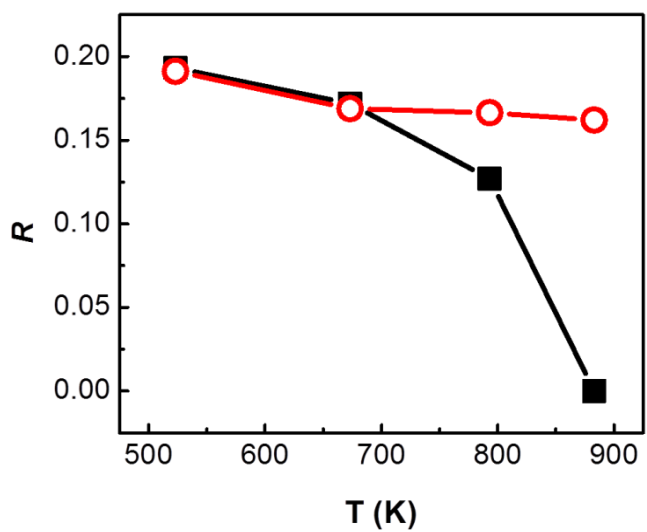

**Figure S4: The RHEED intensity ratio *R* vs. annealing temperature with (black squares) and without (red circles) Te flux.**

Annealing experiments were also performed, both in vacuum and under the supply of Te flux. **Figure S4** presented below summarizes the results at different annealing temperatures while keeping the annealing time fixed at 2 hours. Here, the solid symbols and the black line refer to the data for annealing under vacuum, whereas the open red circles and line are for annealing under the flux of Te. It is seen that increasing annealing temperature leads to a decreasing *R* (corresponding to an increasing 2H domain or a phase transition from the 1T' to 2H phase). On the other hand, the same annealing but under Te flux effectively suppresses such a phase transition. This experiment strongly suggests the important role of surface Te in stabilizing 1T' phase MoTe$_2$.

## S5 First principles calculation of Te adsorption on graphene-supported MoTe$_2$

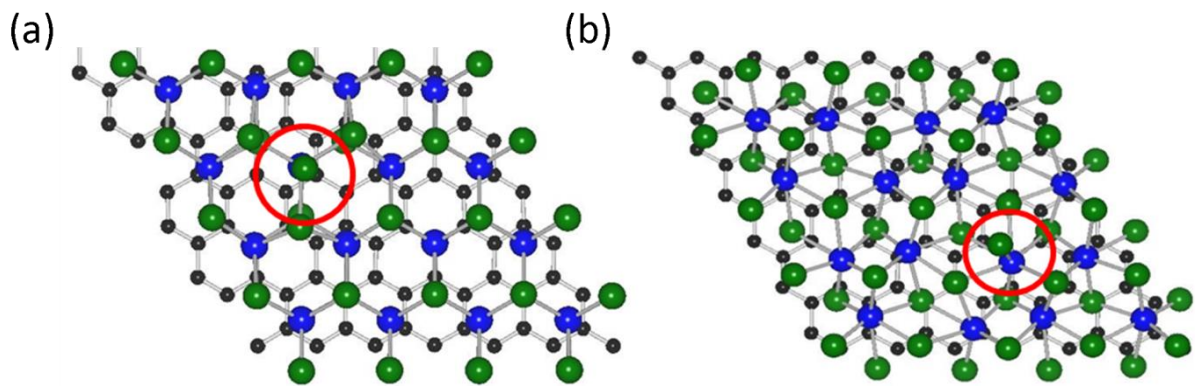

**Figure S5:** Stick-and-ball model showing (a) a Te atom (circled by the red line) adsorbed on 2H MoTe$_2$ surface, and (b) on 1T' MoTe$_2$ surface.

4×4 MoTe$_2$ laid on top of 6×6 graphene units (see Figs. 2a and 3a in the main text and the **Figure S5**, in which the green balls represent Te atoms, blue balls are Mo while the small black balls are carbon atoms in graphene) were chosen to model Te adsorption effect. Upon adsorbing Te atom(s) on the MoTe$_2$ surface (e.g., the one circled by the red line in figure below), different adsorption sites were examined by comparing the energies, based on which we determine the most favorable adsorption site to be the hollow site on surface but above the Mo atom by 2.88 Å for 1T' and 2.97 Å for 2H cases (refer to figure (a) and (b) below, where the red circles mark the adsorbed Te atom n surface). Having determined the adsorption sites, we then put a varying number of Te atoms (1-8) on the 4×4 MoTe$_2$ supercell, all on the hollow sites and arranged symmetrically, to model the varying Te coverage from 6.25% (1/4×4) to 50% (8/4×4). In the calculation, except for graphene substrate, all atoms are allowed to relax. Total energy of the system measured relative to a clean (without Te adsorbate) 2H- MoTe$_2$ is plotted in Figure 4b of the main text.

# Reference

[S1]  M. M. Ugeda, A. J. Bradley, S. F. Shi, F. H. da Jornada, Y. Zhang, D. Y. Qiu, W. Ruan, S. K. Mo, Z. Hussain, Z. X. Shen, F. Wang, S. G. Louie, M. F. Crommie, Nature materials 2014, 13, 1091.